\documentclass[%
reprint,
 amsmath,assume,
 aps,
]{revtex4-2}
\usepackage[bookmarksnumbered,pdfpagelabels=true,plainpages=false,colorlinks=true,linkcolor=blue,citecolor=blue,urlcolor=blue]{hyperref}
\usepackage{graphicx}% Include figure files
\usepackage{dcolumn}% Align table columns on decimal point
\usepackage{amsmath}
\usepackage{amssymb}
\usepackage{amsfonts} 
\usepackage{bm}
\usepackage{bbm}
\usepackage{dsfont}
\usepackage{morefloats}
\usepackage{amsmath,amsfonts,amsthm,amssymb}
\usepackage{appendix}
\usepackage{bbm}
\usepackage{makeidx}
\usepackage{url}
\usepackage{verbatim}
\usepackage[bookmarksnumbered,pdfpagelabels=true,plainpages=false,colorlinks=true,linkcolor=blue,citecolor=blue,urlcolor=blue]{hyperref}
\usepackage{array}
\usepackage{booktabs}
\usepackage{multirow}
\usepackage{bbm}
\usepackage{tabularx}
\usepackage{cancel,soul}
\usepackage{nicefrac, xfrac}
\usepackage{mathrsfs}
\usepackage{mathdots}

\newcommand{\bs}{\boldsymbol{{\rm S}}}

\newcommand{\bH}{\boldsymbol{\mathcal{H}}}

\newcommand{\bP}{\boldsymbol{\mathcal{P}}}
 
\newcommand{\btau}{\boldsymbol{{\rm \tau}}}
\newcommand{\boldeta}{\boldsymbol{{\rm \eta}}}

\newcommand{\bpsi}{\boldsymbol{{\rm \psi}}}
\newcommand{\bPsi}{\boldsymbol{{\rm \Psi}}}
\newcommand{\blambda}{\boldsymbol{{\rm \lambda}}}

\newcommand{\nSites}{\mathcal{{N}}}

\newcommand{\sectioncustom}[1]{\noindent {\it #1.- }}
\def \fcfm {Departamento de Física, FCFM, Universidad de Chile, Santiago, Chile.}
\def \uai {School of Engineering and Sciences, Universidad Adolfo Ibañez, Santiago, Chile}

\def\dfc{Departamento de F\'isica, Facultad de Ciencias, Universidad de Chile, Santiago, Chile}

\begin{document}

\preprint{APS/123-QED}

\title{
%Summoning the toroidon: 
Quantized Toroidal Waves on Ferrotoroidal Magnets}
%\title{Toroidal magnonics of molecular magnets}
\author{Maximiliano Bernal$^1$, Guidobeth Saez$^1$, Tomás P. Espinoza$^2$, Roberto E. Troncoso${}^3$, Alvaro S. Nunez$^1$}
\affiliation{$^1$\fcfm}
\affiliation{$^2$\dfc}
\affiliation{$^3$\uai}
\date{\today}

\begin{abstract}
Magnetic-ferroic ordering and magnetic-toroidal moments are essential concepts in molecular electronics and magnetics. The magnetic toroidal moment is critical in understanding new electronic states and their possible uses. This paper discusses the notion of toroidicity waves. In particular, we present a one-dimensional model of interconnected toroidicity leading to an organization principle around an emergent quantum particle, a carrier of toroidicity waves, dubbed the toroidon. We illustrate some functionalities that could be achieved once control over the toroidon is acquired. We show that a 1D dimerized and antiferromagnetic-like spin chain can display ferrotoroidicity and propose its description in terms of an effective quasi-1D spin chain, marking a crucial step towards further research on the phenomena and potential applications of ferrotoroids.
\end{abstract}

%\keywords{Suggested keywords}%Use showkeys class option if keyword
                              %display tdesired
\maketitle

%\tableofcontents
%\sectioncustom{Introduction} 
{\it Introduction.}-- In light of growing concerns about the rapid production of data and its storage costs, scientific efforts across various disciplines \cite{Lanza2022, Wang2023, Hollenbach2024} focus on shrinking domains in traditional electronic devices to create ultra-dense and cost-effective memories. At such small scales, quantum properties emerge, introducing a range of new physical phenomena that present challenges and offer opportunities for innovative quantum information technologies \cite{Gatteschi2006, Mannini2009}. An example of this is the field of magnonics \cite{Flebus2024}. 
Magnonics is an emerging field within condensed matter physics and nanotechnology that explores the dynamics of spin waves, which are collective excitations of electron spins in a magnetic material. These spin waves propagate through materials in the form of quasiparticles called magnons \cite{Kittel1987, Kittel2004}. It has been shown that, as electrons, magnons can also display features that stimulate the imagination of the scientific community due to their myriad potential applications \cite{Flebus2023, Roldan2017, Doornenbal2019, HidalgoSacoto2020, Pirro2021, Harms2024}.

An additional field where the quantum behavior of matter has been explored is the control of quantum tunneling of magnetization (QTM) \cite{gatteschi2003, gunther2012, JosSantander2015}. The study of QTM turns out to be relevant because, together with other and similar relaxation processes, it has been a persistent obstacle in molecular magnetism, impeding the realization of molecular magnetic memory at room temperature. In contrast, QTM might be advantageous for the construction of quantum devices using molecular nanomagnets as bistable systems that allow linear superposition and coherent manipulation of the quantum states of potential qubits for quantum computing \cite{leuenberger2001, gaita2019}. Before becoming useful, however, the quantum states must show resilience to environmental disturbances that have proven to disrupt their capabilities as information keepers and processing units. Spin-spin and spin-lattice interactions act as sources of decoherence signaling, up to now, the downfall of said devices as viable technologies. 

We will see that the collective state of a handful of moments, given by their toroidal order, offers a pathway toward a compromise amid these tensions. In addition, toroidotronics synthesize magnonics and QTM, opening a complex web of opportunities. This work provides evidence of specific collective behavior that results in magnon degrees of freedom organized around ferrotoroidal systems. This excitation is a toroidal wave, and its quanta are called the toroidon.

Two types of magnetic-ferroic ordering are recognized when categorizing whether they break spatial inversion and time-reversal symmetries. Ferromagnetic ordering disrupts time-reversal symmetry but preserves spatial inversion, whereas magnetic toroidic ordering disrupts both \cite{Xu2024}. The magnetic toroidal moment is a characteristic of specific physical systems that demonstrates a distinct circulation of magnetization lines, forming a vortex-like shape. This phenomenon occurs because of closed magnetic field loops within the system, which differs from the conventional magnetic dipole moment resulting from a pair of magnetic poles. The magnetic toroidal moment is crucial in various fields of physics, such as exploring exotic materials and topological insulators, where it significantly contributes to understanding new electronic states and their possible uses.

A simple way to visualize a toroidal moment is to imagine a magnetic vortex \cite{Hymas2022} where the spins are arranged head-to-tail. However, the forms of toroidal moments can differ in accurate crystalline materials with various symmetries \cite{Ederer2007}. Numerous studies have explored the definition of toroidal moments in bulk crystalline materials. Despite this, some critical questions remain underexplored, such as the symmetry requirements for a spin texture that produces a net toroidal moment, methods for inducing a toroidal moment via external disturbances like magnetic or electric fields or different strains, and the physical phenomena associated with a magnetic toroidal moment. In \cite{Xu2023}, it was shown experimentally that materials with R-3 lattice symmetry can exhibit substantial magnetic toroidicity, even parallel to aligned spins.

Ferrotoroidal order, which involves a natural alignment of toroidal moments, has been observed in a limited number of linear magnetoelectric materials, where adjusting these moments proves difficult.
Recently \cite{Ding2021}, it has been shown to switch between ferromagnetic and ferrotoroidal phases using a minor magnetic field in the chiral triangular-lattice magnet BaCoSiO$_4$, which features 3-spin vortices. Applying a magnetic field induces metamagnetic transitions marked by consistent steps in both magnetic and toroidal moments. Our primary model draws elements from the conceptual picture drawn in \cite{Ding2021} to build a one-dimensional model of interconnected toroidicity. 

%\sectioncustom{Basic Model of a Spin Toroidal Element}
{\it Basic Model of a Spin Toroidal Element.}--
In our model, the magnetic dipoles are aligned in a closed loop, creating a toroidal magnetic moment without a monopole moment. This configuration minimizes the magnetic stray field, making it an area of interest for applications that require low magnetic interference. Modeling of such structures involves quantum mechanics and electromagnetic theory to predict and analyze the spin interactions and magnetic properties at the molecular level. The Hamiltonian $\bH_\tau$ for a spin toroidal molecule can be expressed \cite{Murray2022} as,
\begin{eqnarray}
    \bH_\tau &=& -J \sum_{\langle i,j \rangle} \bs_i \cdot \bs_j-K_t\sum_i (\bs_i\cdot\mathbf{n}_i)^2 +K_z\sum_i (\bs_i\cdot\mathbf{z})^2 \nonumber\\&+& D \sum_i\cdot (\bs_i \times \bs_{i+1})_z - g \mu_B \sum_i \bs_i \cdot \mathbf{B},
\end{eqnarray}
where $i=1,...,\nSites$ spins sites of molecule, $J$ represents the exchange interaction strength between nearest-neighbor spins $\bs_i$ and $\bs_j$, $\mathbf{n}_i=(-\sin\varphi_i,\cos\varphi_i,0)$, with $\varphi_i = \delta\varphi\, i$, is the tangential direction of anisotropy, where $\delta\varphi={2\pi}/{\nSites}$, and
$D$ is the Dzyaloshinskii-Moriya interaction strength with DMI vector along  $z$-axis \cite{Ortix2023}. This interaction is relevant for systems with spin-orbit coupling.
$\mu_B$ is the Bohr magneton,
$\mathbf{B}$ is an external magnetic field. The case with $K=K_t=\mathbf{B}=0$ reduces to the Hamiltonian studied in \cite{Crabtree2018}, which admits an exact diagonalization with a base of states with a well-defined toroidicity.

\begin{figure}[h]
    %\centering
    \includegraphics[width=\columnwidth]{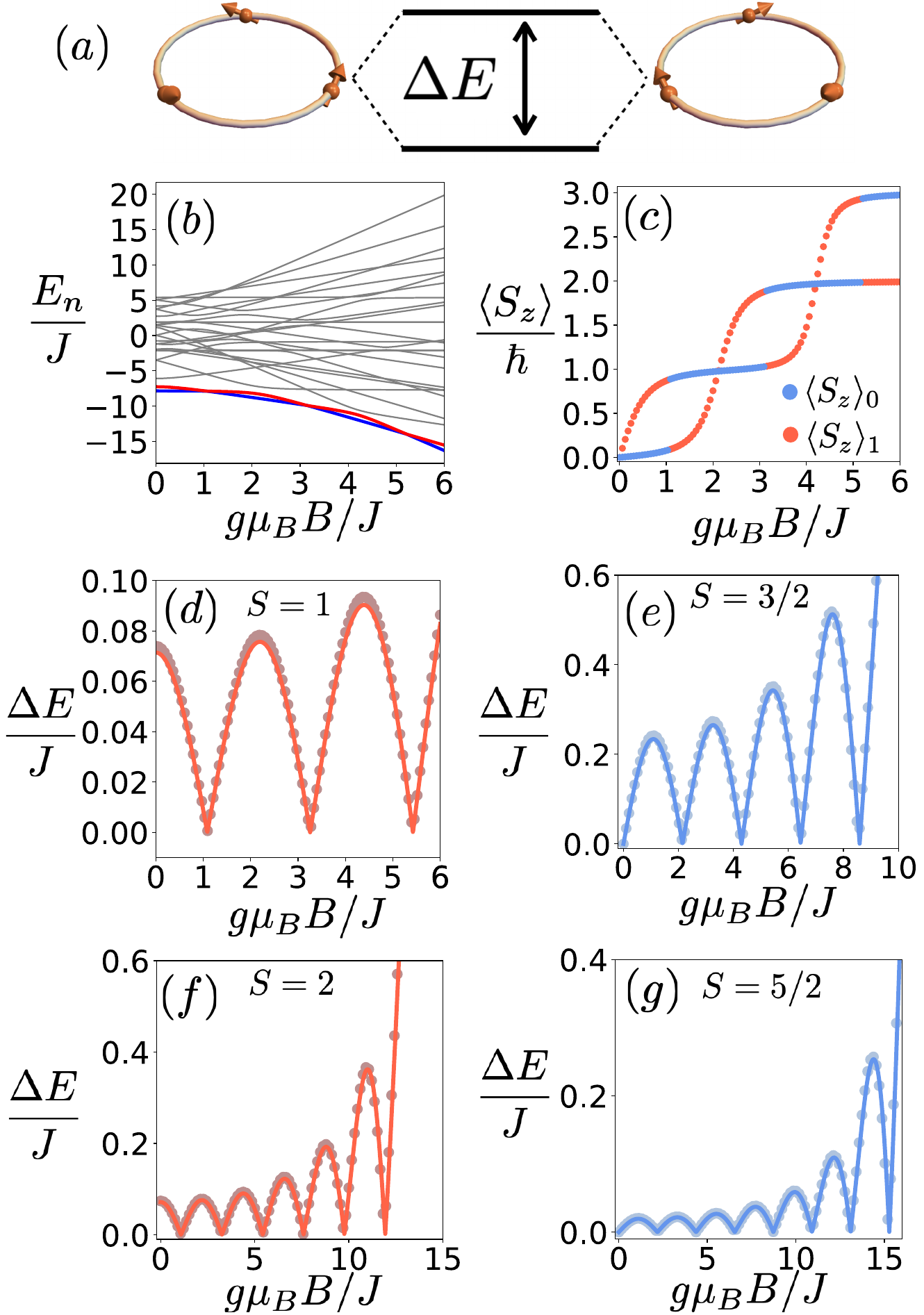}
    \caption{$(a)$ Lowest lying states, displaying well-defined toroidicity, have their classical degeneracy split by quantum fluctuations at generic configurations of the system parameters. This makes the toroidal moment tunnel between classical admissible states. The calculations in the figures correspond to the antiferromagnetic case with three spin and parameters $J=-0.1$, $D/|J|=-20$, $A/|J|=10$, $K/|J|=10$ and magnetic field along $z-$axis. $(b)$ Dispersion of system with spin $S=1$, with three ($N_q = \text{int}(3S)$) cross energy $\Delta E = 0$. $(c)$ Magnetization of trimmer along $z-$axis. Gap energy $\Delta E/J$ between lowest state for different spin values $(d)\,S=1,\,(e)\,S=3/2,\,(f)\,S=2$, and $(g)\,S=5/2$. Points represent the results of exact solutions and solid lines solutions obtained from the effective model of reduced dimension.}
    \label{fig:1}
\end{figure}

Toroidicity tunneling in molecular magnets is a process in which the magnetic state configured in a net toroidal moment, $|\circlearrowleft\rangle$, undergoes quantum tunneling into a reversed state of negative toroidicity, $|\circlearrowright\rangle$. This could involve the collective magnetic moments within a molecular magnet arranged toroidally, tunneling between different toroidal configurations or states. This phenomenon would be quantum mechanical and influenced by the molecule's symmetry, magnetic interactions, and external magnetic fields. As usual, the above leads to oscillations with period $\hbar/\Delta E$, which defines an energy scale $\Delta E$, for the splitting between even and odd combinations, $|\Psi_0\rangle=(|\circlearrowleft\rangle+|\circlearrowright\rangle)/\sqrt{2}$ and $|\Psi_1\rangle=(|\circlearrowleft\rangle-|\circlearrowright\rangle)/\sqrt{2}$, (see Fig. \ref{fig:1}a). Only when $\Delta E=0$ are the tunneling oscillations quenched and the lowest-lying states $|\Psi_0\rangle$ and $|\Psi_1\rangle$ degenerate. This type of behavior can be theoretically studied or observed in systems where the magnetic interactions and quantum states are susceptible to the system's geometrical configuration. This could lead to novel properties relevant to quantum information technologies or other molecular electronics and magnetics applications.

To study the physical properties of the two lowest energy states of the system, we used and compared two approaches. The first corresponds to an exact numerical diagonalization of the spin Hamiltonian $\bH_\tau$ of dimension $(2S+1)^\nSites$. The second consists of creating an effective spin moment $\nSites S$ tuned to render the same low-energy physics as the full system. The effective Hamiltonian $\bH_{e}$, of dimension $2\nSites S+1$, was derived through the spin coherent states path integral \cite{Auerbach1994} representation of the tunneling process \cite{Coleman1985}. We supplement such a description with the collective variable approach, where the freedom of the spin coherent states $|\Omega(\boldsymbol{\theta}, \boldsymbol{\phi})\rangle$ is restricted to satisfy $\theta_i = \theta, \phi_i = \phi - \delta\varphi\, i$, the Hamiltonian expectation value for the coherent state is the energy $E(\boldsymbol{\theta}, \boldsymbol{\phi}) = \langle \Omega | \bH_{\tau} | \Omega \rangle$, and the contribution of each spin to the Berry phase can be approximated as a grand spin $S_{\rm eff} = S\nSites$. The effective action can be cast using an effective Hamiltonian for a single site spin with $S_{\rm eff}$ given by $\bH_{e} = -\kappa_y \bs_y^2 + \kappa_z \bs_z^2 - B \bs_z + E_0$. The effective coupling constants, in the limit of large $S$, are $\nSites\kappa_y = K_t $ and    
$\nSites\kappa_z = K_z +J(\cos  \delta\varphi  -1) +D\sin\delta\varphi$.
In the appendix, we provide the exact values for arbitrary $S$. The toroidicity of the system, using the convention of \cite{Murray2022}, can also be represented using the effective spin. It is a direct calculation to prove that $\btau_z=\nu \, \bs_y$, where $\nu =g \mu_B R$ and $R$ is the radius of the circumcircle of the trimer. This is an essential result for what follows. 

The effective Hamiltonian has hard $(z)$, medium $(x)$, and easy $(y)$ axes and was proposed in \cite{Garg1993} to represent topologically quenched tunnel splitting in spin systems. Fig. \ref{fig:1} illustrates the energy gaps $\Delta E$ using these two methods. The quantization axis is along $y$. The toroidicity tunneling is associated with the tunneling of the spin orientation between $\pm \hat{\rm y}$. The quenching corresponds to parameters where both toroidicities's energies become degenerate. The agreement between the two approaches is excellent.

\begin{figure}[h!]
    \centering
    \includegraphics[width=1\linewidth]{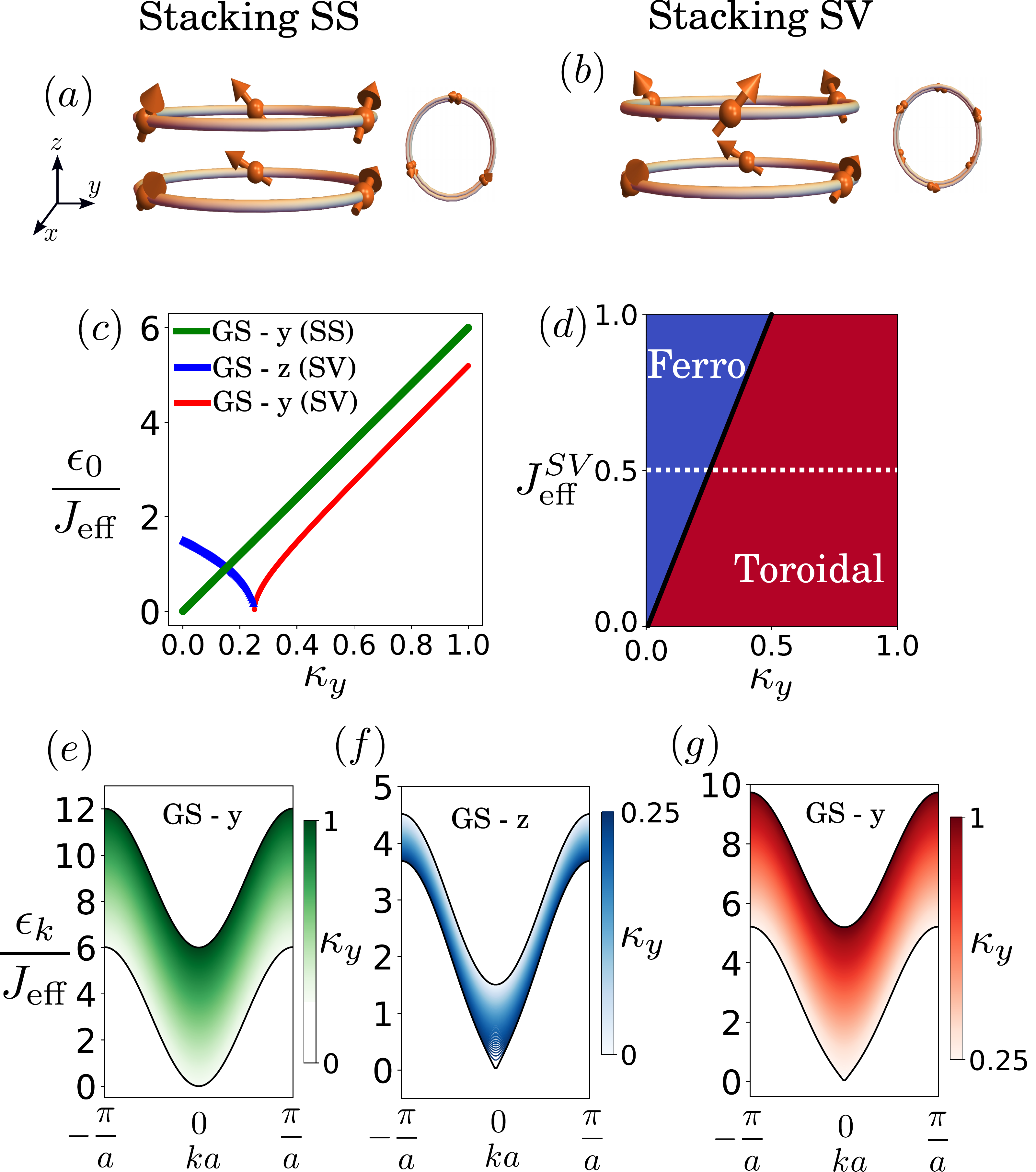}
    \caption{(a) Spin-spin (SS) and (b) spin vacancy (SV) stacking configurations for the toroidal array ($N=3$). (c) Energy \(\epsilon_0\) for the band minimum at \(k = 0\), as a function of anisotropy \(\kappa_y\) for SS and SV stacking configurations, with inter-exchange \(J^{SV}_{\rm eff} = 0.5\) and effective spin $S=3$. (d) Phase diagram of the ground state in the SV stacking configuration, showing the dependence on inter-exchange coupling \(J^{SV}_{\rm eff}\) and anisotropy \(\kappa_y\) (\(\kappa_z = 0\)). The segmented line represents a parameter path. (e) Energy band structure \(\epsilon_k\) for toroidal modes with SS stacking. Energy band structure \(\epsilon_k\) for toroidal modes with SV stacking, (e) with the ground state along the \(z\)-axis and (f) with the ground state along the \(y\)-axis, with parameter path \(J^{SV}_{\rm eff} = 0.5, \kappa_z = 0\).}
    \label{fig: STACKINGS}
\end{figure}

%\sectioncustom{Spin dynamics on a ferrotoroidal array}
{\it Spin dynamics on a ferrotoroidal array.}-- An array of several coupled toroidal elements typically refers to a configuration in which multiple spin toroidal components are linked or interact. Each toroid in the variety can affect its neighbors' magnetic properties, leading to complex behaviors that depend on the specific arrangement and coupling between the elements. The Hamiltonian can be written as:
$$
\bH_a=\sum_\ell \bH^\ell_\tau+\sum_\ell J_\ell\;\bH^{\ell,\ell+1}_t
$$
where we have defined $J_\ell$ the exchange coupling among the $\ell$ and $\ell+1$ layers. We start with the case $J_\ell=j$, a uniform exchange coupling. $\bH^\ell_\tau$ is the Hamiltonian of the toroidal element $\ell$ with internal spins $\bs^\ell_i$, while $\bH^{\ell,\ell+1}_t$ is a ``transfer Hamiltonian'' between the elements: $\bH^{\ell,\ell+1}_t=\sum_{ij}\boldeta_{ij}\bs^\ell_i\cdot\bs^{\ell+1}_j$. The contact matrix $\boldeta$ depends on stacking the toroidal elements. For spins on top of spin,  stacking SS (see Fig. \ref{fig: STACKINGS}(a)), we have $\boldeta=\mathds{1}$, the $3\times 3$ identity matrix. The other possibility that we will explore is a stacking rotated at $\delta\varphi/2$, where each spin is located on top of a vacancy, stacking SV (see Fig. \ref{fig: STACKINGS}(b)). In this case, we have $\boldeta=(\blambda_1+\blambda_4+\blambda_6)/2$, where $\blambda_i$ stands for the $i$-th Gell-Mann matrix \cite{Georgi2019}. 

The effective Hamiltonian for the toroidal array of this case is written as,
\begin{align}\label{eq: effective Hamiltonian}
    \bH = \sum_\ell- \mathbf{S}_\ell\cdot  \mathbb{J}\cdot \mathbf{S}_{\ell+1} -\kappa_y \bs_{\ell\,y}^2 + \kappa_z \bs_{\ell\,z}^2
\end{align}
Where $\mathbb{J}$ is a diagonal matrix. For SS stacking, the exchange is isotropic $\mathbb{J} = J^{SS}_{\rm eff}\mathds{1}$, where $J^{SS}_{\rm eff}=j S^2\nSites$. Meanwhile, for SV stacking, the exchange is anisotropic along $z$, $\mathbb{J} = J^{SV}_{\rm eff}\text{diag}(\cos(\delta\varphi/2),\cos(\delta\varphi/2),1)$, where $J^{SV}_{\rm eff}=2 j S^2\nSites$. In terms of $J^{SV}_{\rm eff}$ and $\kappa_y$, for the stacking SV, we discover a toroidal phase transition from a ferrotoroic regime at large values of $\kappa_y$ toward a ferromagnetic regime (with anisotropy favoring the $z-$axis) at large values of $J^{SV}_{\rm eff}$. The boundary between the ferromagnetic phase and the toroidal phase is given by the line $J^{SV(cr)}_{\rm eff} = (\kappa_z+\kappa_y)/(1-\cos({\delta\varphi}/{2}))$ as can be observed in Fig. \ref{fig: STACKINGS}(d). For stacking SS, the orientation of the ground state is along the $y-$axis, i.e., ferrotoroidal independently of system interactions.

We now study deviations from the just-mentioned ground states. The ferrotoroidic regime will sustain coherent oscillations of the toroidal moment. We call such oscillations toroidons and are the backbone of the remaining parts of this paper. We address the primary behavior of the toroidons that are effectively described by the $y-$directed ferromagnetic state of the effective Hamiltonian in Eq. (\ref{eq: effective Hamiltonian}), we will use Holstein-Primakoff \cite{Holstein1940} bosons $\mathbf{a}_n^{\dagger}$ and $\mathbf{a}_n$ truncated to the second order.
%\begin{align}
 %   \mathbf{S}_n \cdot \boldsymbol{\Omega} &= S-a_n^\dagger a_n \\ 
  %  S^+_{n} &= \sqrt{S-a_n^{\dagger}a_n}\;a_n\\ 
   % S^-_{n} &= a_n^\dagger\sqrt{S-a_n^{\dagger}a_n}
%\end{align}
The Hamiltonian can be expressed in quadratic form and diagonalized through the Bogoliubov transformation %{$\boldsymbol{\alpha}_k = u_k\, \mathbf{a}_{k}+v_k\,\mathbf{a}_{-k}^{\dagger}$} 
\cite{Bogoljubov1958, Wagner1986}. The effective system can be worked out with one site for a unit cell of nominal length $a$. In the SS stacking, the energy spectrum of toroidons is,
\begin{align}
    \epsilon^{SS,T}_k = S\sqrt{(2 J^{SS}_{\rm eff}  (1-\cos ka)+2 \kappa_y + \kappa_z)^2-\kappa_z^2}
\end{align}
This behavior can be appreciated in Fig. \ref{fig: STACKINGS}(e). For SV stacking, in the ferrotoroidal case,  the dispersion relation reads,
\begin{align}
    \epsilon_k^{SV,T} &= S\sqrt{\left(2 J^{SV}_{\rm eff}(1-\cos \frac{\delta\varphi}{2}\cos ka )-(2\kappa_z +\kappa_y)\right)^2-\kappa_y^2},
\end{align}
as can be observed in Fig. \ref{fig: STACKINGS}(g). The ferromagnetic case dispersion relation, $\epsilon_k^{SV,F}$, is presented in the SM.

\begin{figure}
    \centering
    \includegraphics[width=0.8\linewidth]{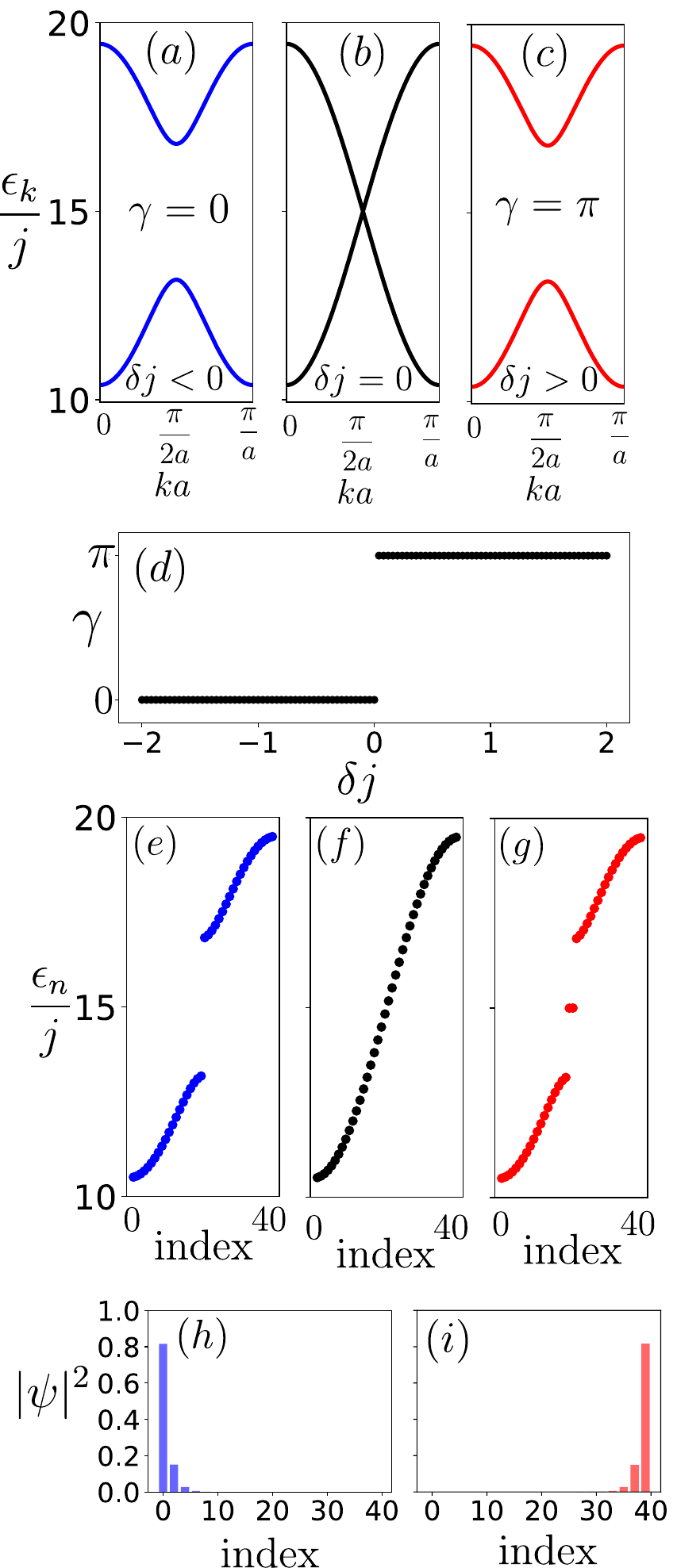}
    \caption{Band structure for toroidal spin waves for dimmerization (a) $\delta j<0$, (b) $\delta j = 0$ and (c) $\delta j >0$, with macroscopic parameters $j=5$, $\kappa_y=10$, $\kappa_z=0$ and effective spin $S=3$. (d) Zak phase for low energy band in function of dimmerization $\delta j$. Energy spectra for finite ferrotoroidal chain width $N=40$ sites, for dimmerization (e) $\delta j<0$, (f) $\delta j = 0$ and (g) $\delta j >0$. Panels (h) and (i), show the wave function of in-gap states.}
    \label{fig: topological}
\end{figure}

%\sectioncustom{Topological toroidons}
{\it Topological toroidons.}-- We now show that toroidal waves can display non-trivial topological behavior. With this in mind, it suffices to adapt the Su-Schrieffer-Heger (SSH) approach to describe polyacetylene topological chains \cite{Su1979, Asbth2016, Moessner2021}. For this purpose, the exchange between layers $J_\ell$, will alternate between two values $J_\ell = j+(-1)^\ell \delta j$. The effective system will reflect this behavior with the reduced Hamiltonian becoming,
\begin{align}
    \bH = &\sum_{\ell}- \mathbf{S}_{\ell} \cdot \mathbb{J}_{\ell}\cdot \mathbf{S}_{\ell+1}-\kappa_y \bs_{\ell\, y}^2 + \kappa_z \bs_{\ell\, z}^2,
\end{align}
where $\mathbb{J}_{\ell}$ is defined as earlier with the corresponding local value of $J_\ell$. With two sites in the unit cell, the Bogoliubov transformation must be handled using the Colpa method \cite{Colpa1986}, a standard method used to treat topological magnons \cite{RoldanMolina2016, Aguilera2020, Jaeschke2021, Tapia2024}.  As expected, the system has two spin-wave modes, gapless for the non-dimerized case $\delta j = 0$ and gapped for the dimerized case $\delta j \neq 0$. A calculation of the winding numbers shows that for $\delta j>0$, the bands are indeed topological. The topological phase diagram is shown in Fig. \ref{fig: topological}.

{\it Discussion.}-- We propose two methods to detect and excite toroidal waves. First, we appeal to the anomalous Hall effect \cite{Nagaosa2010, Nayak2016} in the plane perpendicular to the chain axis. The Hall conductance is proportional to the chirality \cite{Taguchi2004} of the toroidal element and will be affected by the toroidal excitation. Restricting our analysis to the case $\nSites=3$, we can calculate the chirality on every element of the chain by $\chi=\bs_1\cdot(\bs_2\times\bs_3)$ \cite{Li2022}. A straightforward calculation allows us to write $\chi = \eta\, \bs_z\Big( S^2_{\rm eff} -\bs_z^2 \Big)$ in terms of the components of the effective spin where $\eta$ is defined in the Supplemental Material (SM). Hence, the toroidal excitation is closely linked to fluctuations in the local chirality and will affect the Hall conductance. Judiciously located AHE measurements along the chain will report on the spatial structure of the toroidon with information such as its wavelength. 

Second, developed in \cite{Katsura2005}, the Katsura-Nagaosa-Balatsky rule describes how a spiral spin configuration can induce electric polarization in multiferroic materials. The rule states that the electric polarization $\bP$ induced by a spiral spin configuration is given by,
\begin{equation}\label{eq: polarization}
\mathbf{P}_{ij} = \alpha\; \mathbf{e}_{ij} \times (\mathbf{S}_i \times \mathbf{S}_j)
\end{equation}
where $\mathbf{e}_{ij}$ is the unit vector connecting neighboring spins $\mathbf{S}_i$
and $\mathbf{S}_j$ and $\alpha$ is a proportionality constant depending on the spin-orbit coupling within the system. Polarization arises from the inverse Dzyaloshinskii-Moriya (DM) interaction \cite{Tokunaga2023}, an antisymmetric exchange interaction allowed in systems lacking inversion symmetry. The cross product $\mathbf{S}_i\times\mathbf{S}_j$ gives a vector perpendicular to the spin rotation plane, and the direction and magnitude of polarization depend on the sense and angle of the spin spiral. For a review, see \cite{Dong2019}. The rule is particularly useful for predicting and explaining the behavior of certain multiferroics in which electric polarization can be controlled by modifying the magnetic order. This has practical implications in designing devices where electrical control of magnetic properties is desired, such as memory storage devices. Evaluating the polarization from Eq. (\ref{eq: polarization}), we see that every element of the chain has a polarization $\mathbf{P}= \zeta\, \alpha (\bs_x\bs_z+\bs_z\bs_x)\hat{\rm z}$ in terms of the effective spin, where $\zeta$ is defined in the {\rm SM}. This leads to a direct coupling between toroidons and polarization fluctuations.

A toroidal moment, which signifies ferrotoroidic order, arises from a head-to-tail arrangement of magnetic moments. Theoretically, it is suggested that a 1D dimerized and antiferromagnetic-like spin chain can display ferrotoroidicity with a toroidal moment formed by just two opposing spins. In this context, %{\color{red}the researchers present}
Ba$_6$Cr$_2$S$_{10}$ as a potential ferrotoroidic material based on this spin chain model. This structure features distinctive dimerized CrS$_6$ octahedral chains aligned along the c-axis. At around 10 K, an AFM-like order disrupts both spatial and temporal symmetries, and the magnetic point group mm'2' permits three types of ferroic orders in Ba$_6$Cr$_2$S$_{10}$: (anti)ferromagnetic, ferroelectric, and ferrotoroidic. The study indicates that Ba$_6$Cr$_2$S$_{10}$ is an exceptional candidate for ferrotoroidicity with its quasi-1D spin chain, marking a crucial step towards further research on the phenomena and potential applications of ferrotoroidicity \cite{Zhang2022}.

A final remark is that the one-dimensional geometry we discuss in this paper can readily be generalized to more complex structures where the toroidal waves play a significant role in the magnonic dynamics. Even two-dimensional systems, such as a suitably decorated Kagome lattice, could be treated within the perspective presented here.

In this paper, the notion of toroidicity waves is discussed. We present a one-dimensional model of interconnected toroidicity leading to an organization principle around an emergent quantum particle, a carrier of the toroidon. We illustrate some functionalities that could be achieved once control over the toroidon is acquired. We show that a 1D dimerized and antiferromagnetic-like spin chain can display ferrotoroidicicity and propose its description in terms of an effective quasi-1D spin chain, marking a crucial step towards further research on the phenomena and potential applications of ferrotoroids.

{\it Acknowledgements.}-- Funding is acknowledged from Fondecyt Regular 1230515, 1230747. G.S. thanks to the financial support ANID Subdirección de Capital Humano/Doctorado, Chile Nacional/2022-21222167 provided. %Powered@NLHPC: This research was partially supported by the supercomputing infrastructure of the NLHPC (CCSS210001).

\bibliography{toroidal}
\end{document}